\documentclass[twocolumn,showpacs,prl]{revtex4}
\usepackage{graphicx}% Include figure files
\usepackage{bm}% bold math
\usepackage{psfig}
\usepackage{amsmath}
\usepackage{amssymb}
\begin{document}

\title{Correlated States of Electrons in Wide Quantum Wells at Low Fillings: The Role of Charge Distribution Symmetry}
\author{J.~Shabani, T.~Gokmen and M.~Shayegan}
\affiliation{Department of Electrical Engineering, Princeton University, Princeton, NJ 08544, USA}
\date{\today}
\begin{abstract}
Magneto-transport measurements on electrons confined to a 57 nm-wide, GaAs quantum well reveal that the correlated electron states at low Landau level fillings ($\nu$) display a  remarkable dependence on the symmetry of the electron charge distribution.  At a density of $1.93 \times 10^{11}$ cm$^{-2}$, a developing fractional quantum Hall state is observed at the even-denominator filling $\nu = 1/4$ when the distribution is symmetric, but it quickly vanishes when the distribution is made asymmetric. At lower densities, as we make the charge distribution asymmetric, we observe a rapid strengthening of the insulating phases that surround the $\nu = 1/5$ fractional quantum Hall state.
\end{abstract}


\maketitle

Low disorder two-dimensional (2D) electron systems (ESs) at high magnetic fields ($B$) have provided one of the richest grounds to study the physics of interacting charged particles \cite{Shayegan05}.  Much of the work has been done in 2D ESs confined to modulation-doped GaAs/AlGaAs heterostructures where the electrons are separated from the ionized impurities to minimize the scattering and disorder. Recently, it has been recognized that 2D ESs of the highest quality can be realized in modulation-doped wide quantum well (WQW) GaAs samples of width $\geq 30$ nm \cite{Pan02,Eisenstein02,Xia04,Dolev08}. These samples have led to the observation of some of the most spectacular fractional quantum Hall state (FQHS) phenomena and reentrant insulating phases (IPs) at very low Landau level (LL) filling factors ($\nu$) as well as in the higher LLs ($\nu > 2$). Most recently, a new FQHS at the {\it even-denominator} filling $\nu=1/4$ was reported in a 50 nm-wide GaAs WQW at very high $B$ \cite{Luhman08}.

Here we present magneto-transport measurements on 2D ESs confined to a 57 nm-wide GaAs WQW. We employ back- and front-gate electrodes to control the electron density, $n$, as well as the symmetry of the charge distribution. Our measurements reveal that this symmetry plays a crucial role in stabilizing the correlated states of 2D electrons at low $\nu$. We find that the recently observed FQHS at $\nu=1/4$ quickly disappears when the charge distribution is made asymmetric, suggesting that the origin of this state is similar to the $\nu = 1/2$ FQHS observed in WQWs \cite{Suen92,Suen94}. At lower $n$, we observe a very strong FQHS at $\nu = 1/5$. As commonly observed, the 1/5 state is flanked by IPs at nearby $\nu$; these IPs are generally believed to be signatures of pinned electron Wigner solid states. The IPs in our WQW, however, have a surprisingly small resistivity, only about 50 k$\Omega$/$\square$ at a temperature ($T$) of 35 mK, and a weak $T$-dependence when the charge distribution is symmetric. Remarkably, when we make the charge distribution asymmetric, the resistivity of the IPs increases by more than a factor of twenty at 35 mK and shows a strong $T$-dependence, while the resistivity at lower $B$ barely changes.

Our structure was grown by molecular ebeam epitaxy and consists of a 57 nm-wide GaAs WQW bounded on each side by an $\simeq$ 130 nm-thick undoped AlGaAs spacer layer. The WQW is modulation doped symmetrically with Si $\delta$-layers. The mobility of our sample is $\mu=2.5 \times 10^{6}$ cm$^{2}$/Vs at $n=1.93 \times 10^{11}$ cm$^{-2}$. A Ti/Au front-gate evaporated on the surface and a Ga back-side gate were used to change the density of the 2D ES and control the symmetry of its charge distribution. The longitudinal and transverse resistivities, $\rho_{xx}$ and $\rho_{xy}$, were measured in a van der Pauw square geometry. The data were taken in a $^{3}$He/$^{4}$He dilution refrigerator with a base $T$ = 35 mK in a 35 T magnet. For electrical measurements we used the lock-in technique at a frequency of 5.66 Hz with a sample excitation current of 1-10 nA.

\begin{figure*}[ht!]
\centering
\includegraphics[scale=0.9]{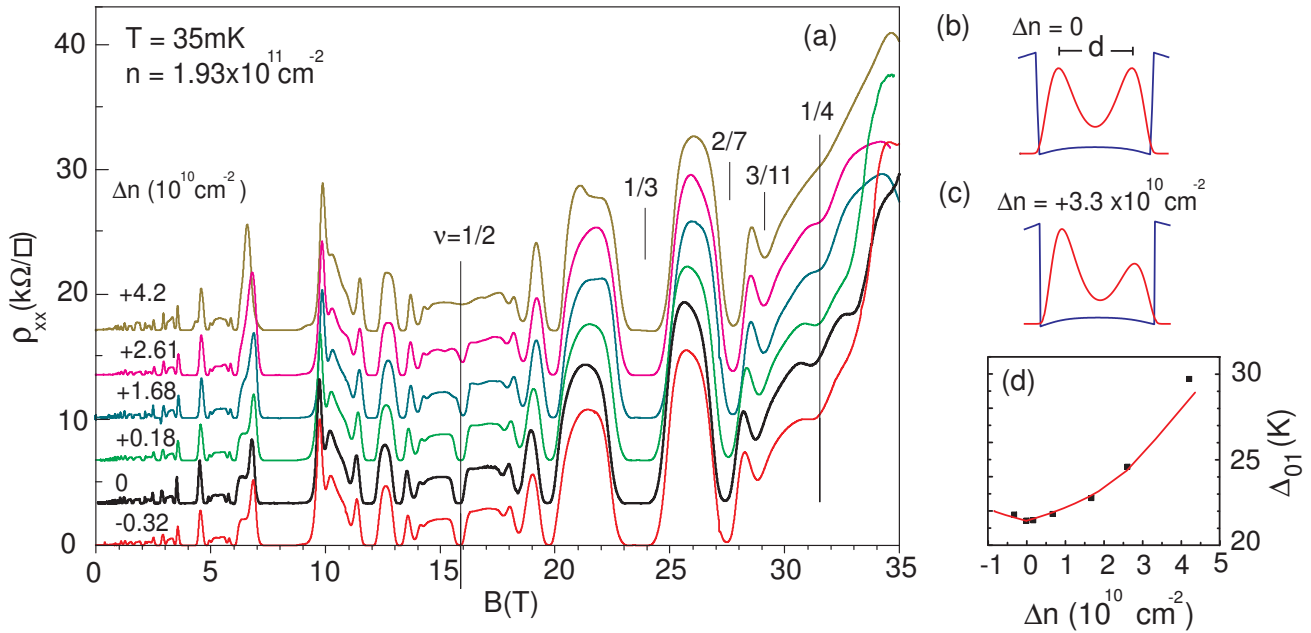}
\caption{(Color online) (a) Evolution of FQHSs for a 57 nm-wide GaAs quantum well with $n=1.93 \times 10^{11}$ cm$^{-2}$ as the charge distribution is made asymmetric. (b) and (c) Self consistently calculated electron distributions and potentials for  $\Delta n=0$ and $\Delta n=3.3 \times 10^{10}$ cm$^{-2}$. (d) Experimentally measured (black squares) and calculated (red curve) subband energy difference ($\Delta_{01}$) as a function of $\Delta n$. }
\end{figure*}

When electrons at very low $n$ are confined to a modulation-doped WQW, they occupy the lowest electric subband and have a single-layer-like (but rather thick in the growth direction) charge distribution. As more electrons are added to the well, their electrostatic repulsion forces them to pile up near the well's walls and the charge distribution appears increasingly bilayer-like \cite{Suen91,Suen92,Suen94,Shayegan96,Manoharan96,Shayegan99}. At high $n$ the electrons typically occupy the lowest two, symmetric and antisymmetric, electric subbands which are separated in energy by $\Delta_{SAS}$. An example of the charge distribution in such a system is given in Fig.~1(b) where we show the results of our self-consistent calculations for $n=1.93 \times 10^{11}$ cm$^{-2}$ electrons symmetrically distributed in our 57 nm-wide WQW. A crucial property of the ES in a WQW is that both $\Delta_{SAS}$ and $d$ (the inter-layer separation), which characterize the coupling between the layers, depend on $n$: increasing $n$ makes $d$ larger and $\Delta_{SAS}$ smaller so that the system can essentially be tuned from a (thick) single-layer like ES at low $n$ to a bilayer one by increasing $n$. This evolution with density plays a decisive role in the properties of the correlated electron states in this system. Equally important is the symmetry of the charge distribution in the WQW. For a fixed $n$, as the distribution is made asymmetric, the separation $\Delta_{01}$ between the lowest two energy levels becomes larger than $\Delta_{SAS}$ and the system becomes increasingly single-layer like. Figure~2(c) shows an example of the calculated charge distribution for the case where $n$ is the same as in Fig.~1(b) but electrons are transferred from the right side of the WQW to the left side so that there is a layer density difference of $\Delta n = 3.3 \times 10^{10}$ cm$^{-2}$. As seen in Fig.~1(d), $\Delta_{01}$ increases from 21.3 K for $\Delta n=0$ to $\simeq$ 29 K for $\Delta n = 4 \times 10^{10}$ cm$^{-2}$. Again, the symmetry of the charge distribution has a profound effect on the correlated states in a WQW \cite{Suen94,Shayegan96,Manoharan96,Shayegan99}.

Experimentally, we control both $n$ and $\Delta n$ in our sample via front- and back-side gates, and by measuring the occupied subband electron densities from the Fourier transforms of the low-field ($B \leq$ 0.4 T) magneto-resistance oscillations. These Fourier transforms exhibit two peaks whose frequencies are directly proportional to the subband densities. By carefully monitoring the evolution of these frequencies as a function of $n$ and, at a fixed $n$, as a function of the values of the back and front gate biases, we can determine and tune the symmetry of the charge distribution \cite{Suen94,Shayegan96,Manoharan96,Shayegan99}. We show an example of such tuning in Fig. 1(d) where we plot our measured values of $\Delta_{01}$ vs $\Delta n$ \cite{footnote1}. The very good agreement between the calculated and experimental $\Delta_{01}$ attests to our control of the charge distribution in our WQW.

In Fig.~1(a) we show $\rho_{xx}$ data as a function of $B$ for different charge distributions while keeping $n$ constant at $1.93 \times 10^{11}$ cm$^{-2}$. Each trace is for a different $\Delta n$ as indicated on the left. In Fig.~2 we present $\rho_{xx}$ traces at different densities while we keep the charge distribution symmetric. The observation of numerous FQHSs at many LL fillings, including ones at very low $\nu$ such as 1/3, 2/7, 3/11, and 1/5 attests to the very high quality of the sample. Of particular interest here are the $\rho_{xx}$ minima observed at the even-denominator fillings $\nu$ = 1/2 and 1/4. We first discuss the 1/2 state and then return to $\nu=1/4$.

The $\nu=1/2$ FQHS in our WQW is strongest when the charge distribution is symmetric at $n=1.93 \times 10^{11}$ cm$^{-2}$ and becomes weaker if either the charge distribution is made asymmetric (Fig.~1(a)) or $n$ is lowered (Fig.~2). This behavior is consistent with previous experimental studies of the 1/2 FQHS in WQWs \cite{Suen94,Shayegan96,Manoharan96,Shayegan99}, and can be understood by examining the competition between three energies: (i) $\Delta_{SAS}$ (or $\Delta_{01}$), (ii) the in-plane correlation energy $Ce{^2}/l_{B}$ (where $C \simeq$ 0.1 is a constant, and $l_{B} = (\hbar/eB)^{1/2}$ is the magnetic length), and (iii) the inter-layer Coulomb interaction $e^{2}/d$. To quantify the behavior, it is useful to construct the ratios $\alpha = \Delta_{SAS}/(e{^2}/\epsilon l_{B})$ and $(e^{2}/\epsilon l_{B})/ (e^{2} /\epsilon d) = d/l_{B}$. As $n$ increases, $\alpha$ decreases since both $\Delta_{SAS}$ and $l_{B}$ (for a FQHS at a given $\nu$) decrease, and $d/l_{B}$ increases. When $\alpha$ is large, the system should exhibit only "one-component" (1C) FQHSs, i.e., standard, single-layer, odd-denominator states, constructed from only the symmetric subband. For small $\alpha$, the in-plane Coulomb energy becomes sufficiently strong to allow the antisymmetric subband to mix into the correlated ground state to lower its energy and a two-component (2C) state ensues. These 2C states, constructed out of the now nearly degenerate symmetric and antisymmetric basis states, have a generalized Laughlin wavefunction, $\Psi_{mmp}^{\nu}$, where the integer exponents $m$ and $p$ determine the intra-layer and inter-layer correlations, respectively, and the total filling factor for the $\Psi_{mmp}^{\nu}$ state is $\nu = 2/(m+p)$ \cite{Halperin,Yoshioka,MacDonald90,He93}. Detailed studies of the evolution of the $\nu=1/2$ FQHS in WQWs, as a function of both $n$ and $\Delta n$, have revealed that this state is stable in a range of $\alpha$ and $d/l_{B}$ values where the inter-layer and intra-layer correlations have comparable strengths \cite{Suen94,Shayegan96,Manoharan96,Shayegan99}. In particular, the fact that it disappears when the charge distribution is made asymmetric provides very strong support for the 2C origin of this state \cite{footnote2}. There is also theoretical justification that the $\nu=1/2$ state has a 2C origin and that its wavefunction has large overlap with $\Psi_{331}^{1/2}$ \cite{He93}. We emphasize that the parameters for where we observe the $\nu=1/2$ FQHS in our WQW, namely, $\alpha ~= 0.082$ and $d/l_{B} = 6.1$, are consistent with the "map" where this state has been reported to be strong \cite{Suen94,Shayegan96,Manoharan96,Shayegan99}.

A remarkable feature of the data in Fig.~1(a) is the presence of a weak, yet clearly notable minimum in $\rho_{xx}$ at $\nu=1/4$ at $n=1.93 \times 10^{11}$ cm$^{-2}$. It signals a developing FQHS at this filling, confirming the recent observation of this state in a WQW \cite{Luhman08}. In Fig. 3 we show our measured Hall resistivity, $\rho_{xy}$, and its derivative, $d\rho_{xy}$/$dB$, vs $B$. The $\rho_{xy}$ trace we show is the average of the magnitudes of the $\rho_{xy}$ traces taken for positive and negative values of $B$, and shows well-quantized plateaus at many integer and FQHSs, including one at $\nu=1/2$. The $\rho_{xy}$ trace further supports a developing FQHS at $\nu=1/4$: there is an inflection point in $\rho_{xy}$ (see the upper inset in Fig. 3), leading to a clear minimum in $d\rho_{xy}$/$dB$ \cite{footnote3}.

Now from the data of Figs.~1 and 2 it is clear that the $\nu=1/4$ minimum in our WQW disappears as we either lower $n$ or make the charge distribution asymmetric. In fact, the strength of the 1/4 minimum correlates with that of the $\nu=1/2$ FQHS observed in the same traces remarkably well. We conclude that, similar to the $\nu=1/2$ FQHS, the $\nu=1/4$ state also has a 2C origin. There are indeed two candidates for a 2C state at $\nu=1/4$: $\Psi^{1/4}_{553}$ and $\Psi^{1/4}_{771}$ \cite{Halperin,MacDonald90}. We suggest that $\Psi^{1/4}_{553}$ is a stronger candidate since the intra-layer correlations in such a state are similar to a $\nu=1/5$ FQHS, a state which is certainly stronger than one at $\nu=1/7$ in a single-layer 2D ES.
\begin{figure}[h!]
\centering
\includegraphics[scale=0.85]{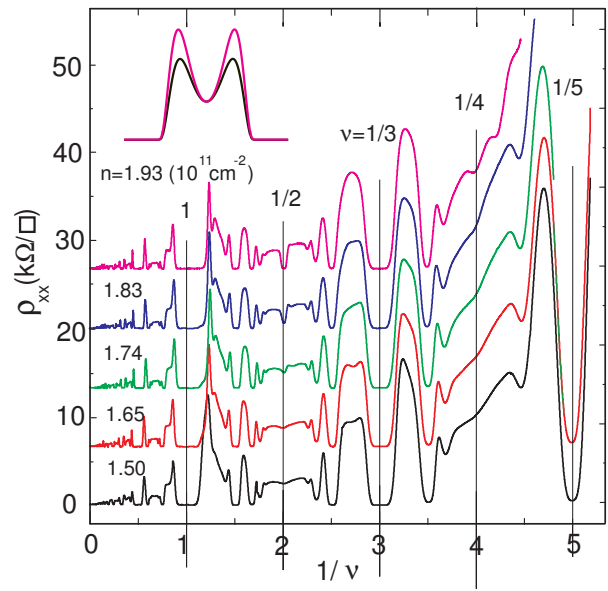}
\caption{(Color online) $\rho_{xx}$ plotted vs inverse filling factor for different densities. Minima at $\nu$ = 1/2 and 1/4 get weaker with lowering the density. The inset shows the calculated charge distribution at $n=1.93 \times 10^{11}$ cm$^{-2}$. (red) and $n=1.50 \times 10^{11}$ cm$^{-2}$ (black).   }
\end{figure}

Next, we present another intriguing feature of an ES confined to a WQW at very small $\nu$. In Fig.~2, as $n$ is lowered, we observe a very strong FQHS at $\nu=1/5$. Note that, as shown in the inset to Fig.~2, even at a density of $n=1.50 \times 10^{11}$ cm$^{-2}$, the charge distribution is quite bilayer-like and yet the ES exhibits a very strong $\nu=1/5$ FQHS, which is a hallmark of very high quality {\it single-layer} ESs in GaAs samples \cite{Jiang90,Goldman90}. Perhaps even more surprising is the fact that the $\rho_{xx}$ maxima observed on the flanks of the $\nu=1/5$ minimum in our WQW are only about 50 k$\Omega/\square$ at our base $T$ of 35 mK, about a factor of twenty smaller than what is typically seen in standard, single-layer 2D ESs at comparable $T$ \cite{Jiang90,Goldman90}.

Figure~4 reveals more details of magneto-transport at very small $\nu$. All the data shown in this figure were taken at $n=1.50 \times 10^{11}$ cm$^{-2}$ as $\Delta n$ was varied \cite{footnote4}. As seen in Fig.~4(a), the $\rho_{xx}$ maxima in our WQW depend very strongly on the symmetry of the charge distribution. For example, $\rho_{xx}$ on the high-$B$ side of $\nu=1/5$ changes from about 50 k$\Omega/\square$ for $\Delta n = 0$ to about 1.5 M$\Omega/\square$ for $\Delta n = 6.3 \times 10^{10} $ cm$^{-2}$. Note that at lower $B$ ($\nu > 1/3$), $\rho_{xx}$ is rather insensitive to charge imbalance, and in fact near $\nu=1/2$, $\rho_{xx}$ decreases as the charge is made asymmetric. Figures 4(b) and (c) reveal that the $T$-dependence of $\rho_{xx}$ also depends on charge distribution symmetry. In both figures $\rho_{xx}$ exhibits IPs on the flanks of $\nu=1/5$ minimum, but the $T$-dependence is much stronger for the asymmetric charge distribution. This is particularly noticeable on high-$B$ flank.
\begin{figure}[h!]
\centering
\includegraphics[scale=0.85]{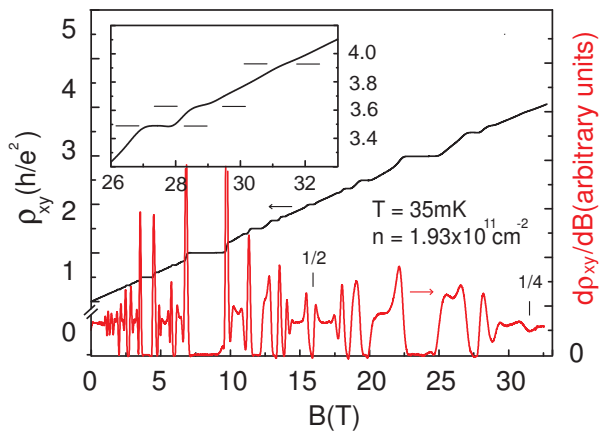}
\caption{(Color online) Hall resistivity and its derivative are plotted as a function of magnetic field.  }
\end{figure}
The IPs surrounding the $\nu=1/5$ FQHS in single-leyer ESs have generally been interpreted as signatures of a {\it pinned} electron Wigner crystal (WC) \cite{Jiang90,Goldman90}. The small but finite, ubiquitous disorder, resulting from the defects in the GaAs/GaAlAs crystal such as residual impurities and interface roughness, leads to potential fluctuations which in turn pin the WC. Such interpretation is validated by numerous microwave conductivity studies \cite{Chen04}, which have measured the various resonance modes of the pinned WC. Our results demonstrate that these IPs are very sensitive to the charge distribution symmetry in the quantum well. It would be very interesting to study the microwave resonance modes in WQWs as a function of this symmetry; such studies are planned.

We would like to emphasize that the results presented here on the IPs in WQWs are complementary to those reported earlier for similar WQWs \cite{Manoharan96}. In Ref. \cite{Manoharan96}, the IPs were observed to move to {\it higher} $\nu$ as density was {\it raised}, eventually surrounding the FQHSs at $\nu$ = 1/3 and 1/2. These IPs were interpreted as signatures of correlated, {\it bilayer} WC states. Consistent with this interpretation, the IPs {\it disappeared} and moved to lower $\nu$ as density was lowered or the charge distribution was made symmetric at a constant density. In the present case, we are probing relatively lower densities (larger subband separation), and we observe the opposite behavior: the IPs appear at very low $\nu$, and they get {\it stronger} when we make the charge distribution asymmetric. It remains to be seen whether this behavior reflects a competition between the FQHS and WC ground state energies of an ES in a WQW with symmetric/asymmetric charge distribution, or it is related to changes in the disorder potential when the charge distribution is made asymmetric.

We thank NSF and DOE for financial support, J-H. Park, T. Murphy, G. Jones and E. Palm for help with the experiments, and L.N. Pfeiffer, E. Tutuc, and K.W. West for illuminating discussions. This work was performed at the National High Magnetic Field Laboratory, which is supported by the NSF (DMR-0654118), the State of Florida, and the DOE.

\begin{figure}[h!]
\centering
\includegraphics[scale=0.85]{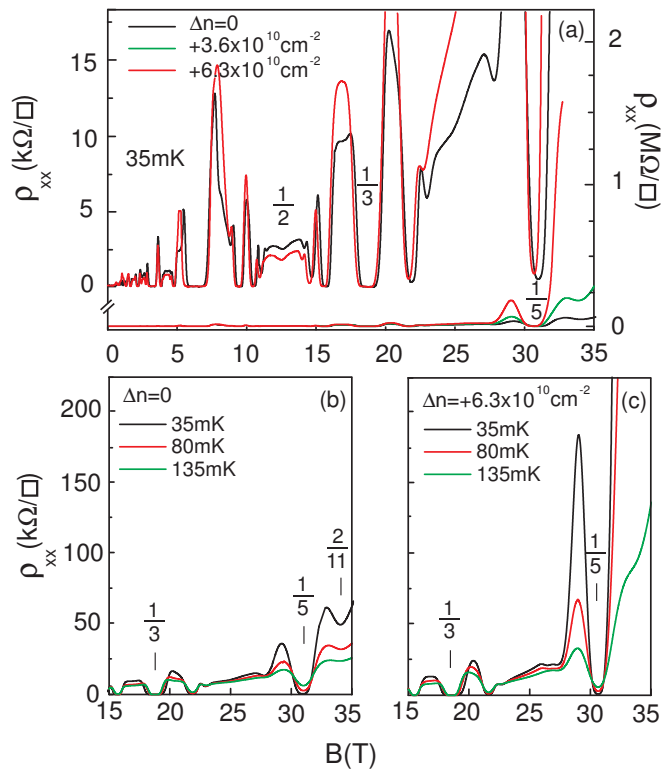}
\caption{(Color online) (a) Resistivity vs magnetic field at $n = 1.50 \times 10^{11}$ cm$^{-2}$ for different $\Delta n$. The upper traces (left scale) are expansions of the lower traces (right scale). (b) and (c) Temperature dependence of $\rho_{xx}$ for $\Delta n = 0$ and $\Delta n = 6.3 \times 10^{10}$ cm$^{-2}$.}
\end{figure}


\begin{thebibliography}{99}
\bibitem{footnote1}
We use data such as those shown in Fig. 1(d) to determine the "balanced" condition for our WQW. As grown, our sample has slightly higher density ($1.2 \times 10^{10}$ cm$^{-2}$) in the back layer. Also, through out this Letter, $\Delta n > 0$ means there is more charge in the front (surface) side of the WQW compared to the back (substrate) side.
\bibitem{footnote3}
The value of $\rho_{xy}$ at the inflection point is slightly smaller than expected (3.9 instead of 4.0 times $h/e^{2}$), similar to what is reported in Ref. [6].
\bibitem{footnote4}
Data taken for $\Delta n < 0$ are qualitatively similar o those shown in Fig.~4(a).
\end{thebibliography}
\end{document}